\documentclass[%
 aip,
 amsmath,amssymb,
 reprint,%
]{revtex4-1}

\usepackage{graphicx}% Include figure files
\usepackage{dcolumn}% Align table columns on decimal point
\usepackage{bm}% bold math
\usepackage[utf8]{inputenc}
\usepackage[T1]{fontenc}
\usepackage[nottoc]{tocbibind}
\usepackage{mathptmx}
\usepackage{etoolbox}
\usepackage{gensymb}
\usepackage{booktabs}
\usepackage{siunitx}

\makeatletter
\def\@email#1#2{%
 \endgroup
 \patchcmd{\titleblock@produce}
  {\frontmatter@RRAPformat}
  {\frontmatter@RRAPformat{\produce@RRAP{*#1\href{mailto:#2}{#2}}}\frontmatter@RRAPformat}
  {}{}
}%
\makeatother
\begin{document}

\preprint{AIP/123-QED}

\title[]{Aging and passivation of magnetic properties in Co/Gd bilayers}
% Force line breaks with \\
\author{Thomas J. Kools}
 \email{t.j.kools@tue.nl}
 \author{Youri L. W. van Hees}
\affiliation{%
Department of Applied Physics, Eindhoven University of Technology \\
P. O. Box 513, 5600 MB Eindhoven, The Netherlands
}%
 \author{Kenneth Poissonnier}
\affiliation{%
Department of Applied Physics, Eindhoven University of Technology \\
P. O. Box 513, 5600 MB Eindhoven, The Netherlands
}%
\author{Pingzhi Li}
\affiliation{%
Department of Applied Physics, Eindhoven University of Technology \\
P. O. Box 513, 5600 MB Eindhoven, The Netherlands
}%
\author{Beatriz Barcones Campo}
\affiliation{%
NanoLab@TU/e, Eindhoven University of Technology \\
P. O. Box 513, 5600 MB Eindhoven, The Netherlands
}%
\author{Marcel A. Verheijen}
\affiliation{%
Department of Applied Physics, Eindhoven University of Technology \\
P. O. Box 513, 5600 MB Eindhoven, The Netherlands
}%
\affiliation{%
Eurofins Materials Science Netherlands BV, 5656 AE Eindhoven, The Netherlands
}%
\author{Bert Koopmans}%
\affiliation{%
Department of Applied Physics, Eindhoven University of Technology \\
P. O. Box 513, 5600 MB Eindhoven, The Netherlands
}%
\author{Reinoud Lavrijsen}
\affiliation{%
Department of Applied Physics, Eindhoven University of Technology \\
P. O. Box 513, 5600 MB Eindhoven, The Netherlands
}%

\date{\today}% It is always \today, today,

\begin{abstract}
Synthetic ferrimagnets based on Co and Gd bear promise for directly bridging the gap between volatile information in the photonic domain and non-volatile information in the magnetic domain, without the need for any intermediary electronic conversion. Specifically, these systems exhibit strong spin-orbit torque effects, fast domain wall motion and single-pulse all-optical switching of the magnetization. An important open challenge to bring these materials to the brink of applications is to achieve long-term stability of their magnetic properties. In this work, we address the time-evolution of the magnetic moment and compensation temperature of magnetron sputter grown Pt/Co/Gd trilayers with various capping layers. Over the course of three months, the net magnetic moment and compensation temperature change significantly, which we attribute to quenching of the Gd magnetization. We identify that intermixing of the capping layer and Gd is primarily responsible for this effect, which can be alleviated by choosing nitrides for capping as long as reduction of nitride to oxide is properly addressed. In short, this work provides an overview of the relevant aging effects that should be taken into account when designing synthetic ferrimagnets based on Co and Gd for spintronic applications.
\end{abstract}

\maketitle

%%START OF MAIN TEXT%%

Ferrimagnetic spintronics is an active contemporary research field \cite{Kim2022, SalaReview2022}. One major motivation for investigating these magnetic systems is to combine the easy characterization and manipulation of the magnetic order of ferromagnets with the fast exchange-driven dynamics and information robustness associated with antiferromagnetic coupling. One particular class of ferrimagnets that is of interest here are the so-called transition metal (TM)-rare earth (RE) ferrimagnets, characterized by a composition consisting of one or more TMs (Co, Fe, Ni) and one or more REs (e.g. Gd, Tb). These systems exhibit various fascinating phenomena which have attracted the attention of the scientific community such as single-pulse all-optical switching of the magnetization,\cite{Stanciu2007,Radu2011, Ostler2012, Mangin2014, Lalieu2017, vanHees2020,Verges2022} effective spin-orbit torque (SOT)-driven manipulation of the magnetic order,\cite{Mishra2017,Ueda2017,Je2018,Sala2022}, the presence and efficient motion of skyrmions,\cite{Wang2022Multi,Quessab2022,Brandao2019, Caretta2018} and exchange torque driven current-induced domain wall motion with velocities over 1000 m/s. \cite{Caretta2018,Cai2020,LiKools2022} The combination of these phenomena in one material system makes them a promising candidate to bridge the gap between photonics and spintronics. \cite{Lalieu2019,Wang2022,LiKools2022,pezeshki2022design}

Within this class of materials, often a distinction is made between alloys, with an approximately uniform distribution of TM and RE elements, and synthetic ferrimagnets consisting of discrete layers of the RE and TM. The latter category has a few distinct advantages in comparison to the alloy systems. The layered structure of these synthetic ferrimagnets allows for easier adaptation to wafer scale production. Also, contrary to alloys, a much wider composition range between the 3d and 4f-metal exhibits single-pulse all-optical switching.\cite{Beens2019Alloy,Beens2019Intermixing} Combined with the increased access to interfacial engineering, this leads to more flexibility and tunability of its magnetic properties. Moreover, the Pt/Co/Gd trilayer displays strong interfacial spintronic effects, such as perpendicular magnetic anisotropy, the spin-Hall effect,\cite{Haazen2013} and the interfacial Dzyaloshinskii–Moriya interaction,\cite{Cao2020} which are essential ingredients for applications based on efficient domain wall motion or SOT-driven manipulation.\cite{Blasing2020,SalaReview2022}

\begin{figure}
\centering
\includegraphics[width=0.48 \textwidth]{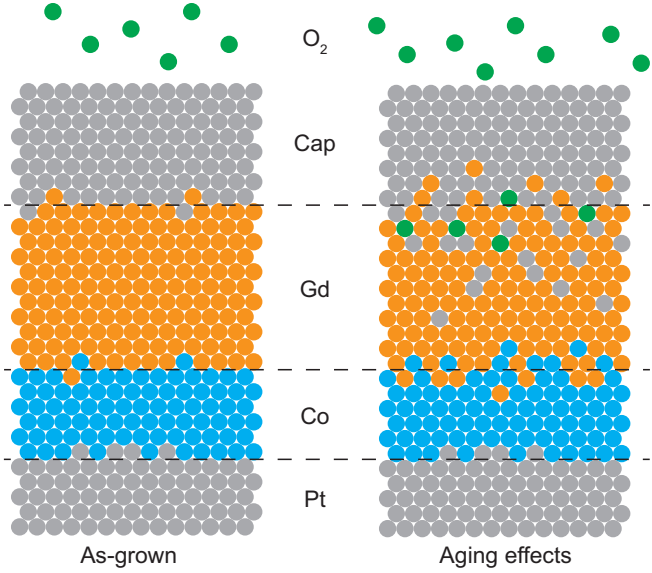}
\caption{\label{fig:Figure1} Schematic illustration of the three main aging effects that can impact the magnetization of the Gd subsystem: intermixing of Gd with the Co layer, intermixing of Gd with the capping layer and oxidation of the Gd layer. The illustration is not to scale, nor an exact representation of the (expected) intermixing profiles.}
\end{figure}

Nonetheless, one important challenge that yet needs to be addressed, before applications of spintronic devices based on synthetic RE-TM systems can be considered, is their long-term stability. The aging and sample structure of RE-TM alloys has been extensively studied in the past, where several factors gave rise to changing magnetic properties over time, like the capping layer,\cite{Taylor1989,Lee2000} compositional inhomogeneities at the scale of several nm,\cite{Graves2013, Liu2015} and thermodynamically driven intermixing.\cite{Taylor1989,Konar2017,Sala2022} Despite this knowledge, aging effects on basic magnetic properties in the ultrathin synthetic ferrimagnets with perpendicular magnetic anisotropy has been poorly studied. Therefore, in this work, we present a systematic VSM-SQUID study of the magnetic moment and compensation temperature of a Co/Gd bilayer over time. Here, based on the earlier work in the alloys, we expect three main contributions to changes in the balance between Co and Gd magnetization, the origin of which is illustrated schematically in Fig. \ref{fig:Figure1}: Intermixing between the capping layer and the Gd, intermixing between the Co and Gd, and oxidation of the Gd via grain boundaries in, or partial oxidation of, the capping layer. Answering the question of which of these contributions is dominant, will be addressed in this work.

\begin{figure}
\centering
\includegraphics[width=0.5 \textwidth]{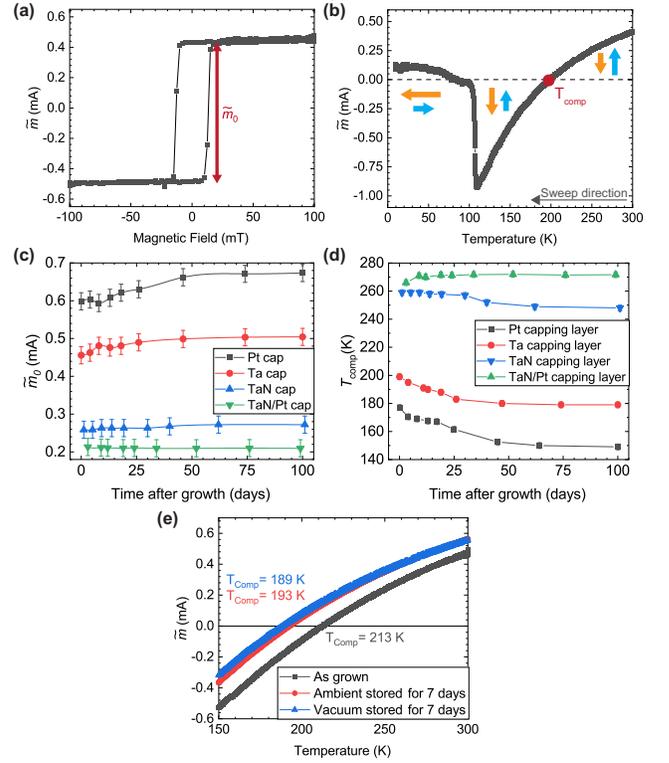}
\caption{\label{fig:Figure2} VSM-SQUID characterization of Ta/Pt/Co/Gd systems. (a) Characteristic OOP hysteresis loop measured at room temperature with VSM-SQUID. (b) Characteristic temperature-dependence of the OOP magnetic moment. Arrows schematically indicate the relative balance between Co (blue) and Gd (orange) magnetization. (c) Time evolution of the OOP moment, as indicated by the gray arrow in (a), for various capping layers. (d) Time evolution of the compensation temperature, as indicated by the gray circle in (b), for various capping layers. Lines are a guide to the eye. (e) Comparison of the OOP magnetic moment as a function of temperature of Pt-capped samples at different time intervals after growth: as grown (black), seven days old stored in ambient conditions (red), and seven days old stored in high vacuum ($\pm$5E-10 mBar, blue).}
\end{figure}

Ta/Pt/Co/Gd/X samples were fabricated by DC magnetron sputtering on thermally oxidized Si/SiO$_\mathrm{x}$ substrates, where X denotes the different capping layers used. The precise layer structure used in this study are summarized in Table \ref{tab:tablemain} (for more details about the sputtering process, see Sup. Mat. 1). These full sheet samples were diced into 4.5×4.5 mm$^2$ samples for SQUID characterization. Out-of-plane (OOP) SQUID measurements as a function of OOP applied field and temperature were performed, using a commercially available  MPMS3 VSM-SQUID system. A typical room-temperature measurement of the area-normalized net OOP magnetic moment as a function of the OOP applied field, is shown in Fig. \ref{fig:Figure2}(a). We choose the area-normalized moment, $\tilde{m}$, as defining the magnetization requires us to define a thickness by which to normalize. However, in that case, coarse assumptions have to be made about the intermixing and magnetization profile in the Gd,\cite{Kools2022} making area normalization the more consistent choice. The resultant square hysteresis loop in $\tilde{m}$ for OOP SQUID indicates perpendicular magnetic anisotropy, as expected for thin films of Co on Pt. \cite{LiKools2022,Lalieu2017} 

\begin{table}
\caption{\label{tab:tablemain} Overview of the stack structures of the samples under investigation in this work.}
\begin{ruledtabular}
\begin{tabular}{ccc}
Substrate [$\SI{}{\micro \metre}$]&Main stack [nm]&Capping layer [nm]\\ \hline
Si(500)/SiO$_\mathrm{x}$(0.1)&Ta(4)/Pt(4)/Co(1)/Gd(3)&Pt(4) \\
Si(500)/SiO$_\mathrm{x}$(0.1)&Ta(4)/Pt(4)/Co(1)/Gd(3)&Ta(4) \\
Si(500)/SiO$_\mathrm{x}$(0.1)&Ta(4)/Pt(4)/Co(1)/Gd(3)&TaN(4) \\
Si(500)/SiO$_\mathrm{x}$(0.1)&Ta(4)/Pt(4)/Co(1)/Gd(3)&TaN(4)/Pt(4) \\

\end{tabular}
\end{ruledtabular}
\end{table}

A typical plot of the temperature dependence of $\tilde{m}$ is shown in Fig. \ref{fig:Figure2}(b). At room temperature, and at these thicknesses of Co and Gd, the magnetization of the Co dominates the magnetic balance.\cite{Lalieu2017} For decreasing temperature, the magnetization of the Gd increases more rapidly than that of the Co, as is typical for the Gd S=7/2 spin system. This manifests in Fig. \ref{fig:Figure2}(b) as the decrease of $\tilde{m}$ with temperature. The temperature at which $\tilde{m}$ crosses zero corresponds to the compensation temperature $T_\mathrm{comp}$, here the moment of the Co and Gd atoms exactly cancel each other ($T_\mathrm{comp}$ $\sim200$ K in Fig. 2(b)). We note that the magnetization compensation and angular momentum compensation point are not the same in these 3d/4f ferrimagnets due to the different Landé g-factor of Co\cite{Scott1962} and Gd\cite{LOW1958315}; in the remainder of this work we will refer to the magnetization compensation only. Finally, at even lower temperature ($\sim105$ K in Fig. 2(b)), $\tilde{m}$ largely vanishes due to a transition from OOP to in-plane magnetization owing to the increased shape anisotropy generated by the Gd magnetization. This overcomes the total perpendicular magnetic anisotropy which mostly originates from the bottom Pt/Co interface. 

\begin{figure*}
\includegraphics[width=1 \textwidth]{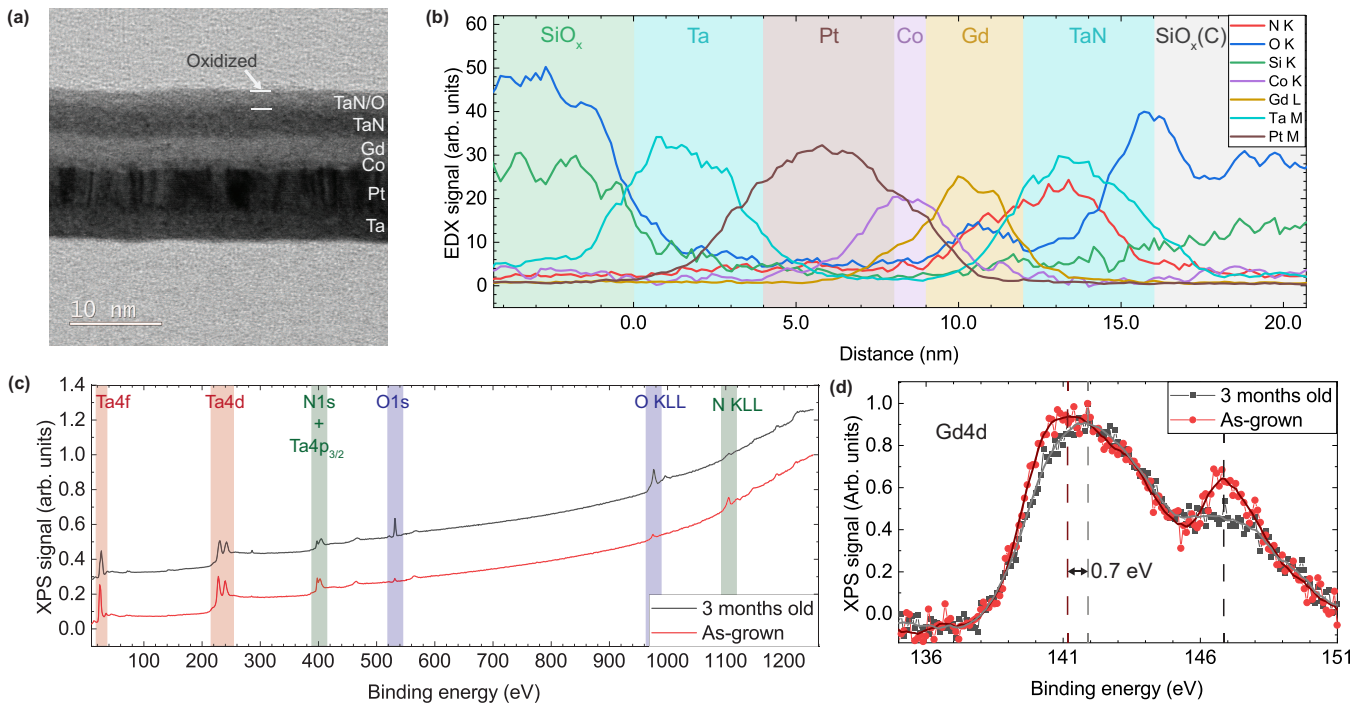}
\caption{\label{fig:3} Chemical analysis of Ta/Pt/Co/Gd/TaN samples (see Table \ref{tab:tablemain}). (a) Bright field scanning transmission electron microscopy image, where the different layers are labeled. (b) Part of the elemental profile of the sample investigated in (a) obtained by EDX. Relative peak intensities do not represent the atomic percentages. Shaded areas indicate the nominal layer thicknesses as listed in table \ref{tab:tablemain} and serve as a guide to the eye. The full profile of this lamella, including the C-contribution and the rest of the substrate and SiO$_\mathrm{x}$(C) layer, can be found in Sup. Mat. 4.  (c) Comparison of XPS survey spectra for an as-grown and 3-month-old sample. (d) Comparison of the Gd4d photoelectron peak for an as-grown and 3-month-old sample. Dashed lines indicate characterisitc changes in the spectrum.}
\end{figure*}

 These hysteresis loops and temperature sweeps were repeated over an extended period of time for two different capping layers which are typically used to prevent oxidation: X={Pt(4), Ta(4)}. From each measurement, the remanent area-normalized net OOP magnetic moment at room temperature, $\tilde{m}_\mathrm{0}$, and $T_\mathrm{comp}$ were extracted as indicated by the red dots in Fig. \ref{fig:Figure2} (a) and (b), respectively. We plot the temporal evolution of $\tilde{m}_\mathrm{0}$ and $T_\mathrm{Comp}$ in Fig. \ref{fig:Figure2}(c) and (d), respectively. For these monatomic capping layers, we observe an increase (decrease) of $\tilde{m}_\mathrm{0}$ ($T_\mathrm{Comp}$) with time, which indicates a decrease of the Gd magnetic moment, since $\tilde{m}_\mathrm{0}$ is dominated by the Co at room temperature. We hypothesize this to be a consequence of the intermixing between the capping layer and Gd, something that is known to occur for several metallic capping layers on RE/TM alloys.\cite{Taylor1989} As the capping layer intermixes with the Gd, the induced magnetization in the Gd is quenched, causing the lower temperature needed to raise the Gd-magnetization to compensate that of the Co. The fact that this effect occurs more severely for Pt than for Ta may be a consequence of the larger number of intermetallic compounds between Pt and Gd  vs. Ta and Gd, 6 vs. 0, respectively,\cite{massalski1986binary} which is typically assumed to impact the species mutual solid solubility with their alloy components. We have repeated this study on a second identically prepared batch of samples, which confirms these trends (see Sup. Mat. 2 for details).
 
 We recognize that the increase (decrease) in $\tilde{m}_\mathrm{0}$ ($T_\mathrm{Comp}$) shown in Fig. \ref{fig:Figure2} (c) and (d) may also be explained by oxidation of the Gd. To investigate the relevant importance of intermixing and oxidation during the aging process, we investigated the temperature dependence of $\tilde{m}$ of two identical samples of the same structure as those listed in Table \ref{tab:tablemain} with a Pt(4) capping layer. These were stored in vacuum ($\sim5\times10^{-9}$ mBar) and under ambient conditions, and compared to the OOP SQUID measurements on the as-grown sample. The resulting data is plotted in Fig. \ref{fig:Figure2}(e). It can be seen that the decrease of $T_\mathrm{Comp}$ occurs irrespective of whether the sample has been exposed to ambient conditions or not. This suggests that the dominant mechanism for the changes in magnetic properties at these early stages of the aging process is indeed magnetization quenching by intermixing between the Gd and the monatomic capping layer.

To further test the hypothesis of the capping layer intermixing with the Gd layer, we move from a monatomic capping layer to the ceramic capping layer TaN, deposited by reactive sputtering from a pure Ta target in a mixed Ar:N$_2$ environment. Nitrides like SiN$_\mathrm{x}$ and AlN$_\mathrm{x}$ are often used for the capping of RE-TM alloys, as they form a covalently bonded compound, rendering them very stable towards interdiffusion.\cite{Lee2000,Cai2020} When inspecting Fig. \ref{fig:Figure2}(c) and (d), we first note that a significantly decreased (increased) $\tilde{m}_\mathrm{0}$ ($T_\mathrm{Comp}$) is observed for both the as-deposited and aged samples capped with TaN, which is consistent with earlier observations.\cite{Kools2022} The origin of this change has not been unequivocally established. One explanation could thus be that intermixing processes between Gd and capping layer play a smaller role for the ceramic TaN than for the monatomic capping layers, leading to a reduced quenching of the Gd magnetization. However, we also note that due to exposure to the N$_2$ gas during the reactive sputtering of TaN, formation of the ferromagnetic semiconductor GdN\cite{Schumacher1995} may also play a role here. The full characterization of the role potential Gd nitrification plays in the magnetic balance of this system would however require high-resolution depth-resolved magnetometry experiments (e.g. neutron scattering), which are beyond the current scope of this manuscript.

When inspecting the time evolution of $\tilde{m}_\mathrm{0}$ and $T_\mathrm{Comp}$ for the sample capped with TaN(4) in Fig. \ref{fig:Figure1}(c) and (d), we find that $\tilde{m}_\mathrm{0}$ and $T_\mathrm{comp}$ are passivated effectively for the first $\sim$35 days, after which $\tilde{m}_\mathrm{0}$ ($T_\mathrm{Comp}$) start to increase (decrease) similarly to the monatomic capping layers, which we hypothesize is due to (partial) oxidation of the TaN under ambient conditions, as reported in earlier work.\cite{Guan2021} To verify this, scanning transmission electron microscopy (STEM) and energy dispersive x-ray spectroscopy (EDX) measurements were performed on a sample with the same structure as the ones characterized with SQUID and a capping layer of TaN (See Sup. Mat. 3 for experimental details). A bright field STEM image is presented in Fig. \ref{fig:3}(a), where a contrast shift of the TaN at the TaN/air interface is observed. In the elemental depth profile obtained from the EDX characterization shown in Fig. \ref{fig:3}(b), we find that this contrast shift indeed follows from a reduction of TaN to TaO$_\mathrm{x}$. 

This observation is corroborated by X-ray photoelectron spectroscopy measurements (XPS, see Sup. Mat. 5 for experimental details). Fig. \ref{fig:3} (c) shows a significantly increased (decreased) oxygen (nitrogen) contribution to the spectrum when comparing the spectrum of a 3-month-old sample to an as-grown sample, for samples with a TaN cap as listed in Table \ref{tab:tablemain}. The gradual replacement of the TaN with TaO$_\mathrm{x}$ could also explain the eventual disappearance of the passivating properties of TaN as a capping layer, as the oxygen in the TaO$_\mathrm{x}$ reacts with the Gd or facilitates diffusion of oxygen to the Gd. The oxidation of the Gd can be investigated by comparison of the main 4d-photoemission peak complex of Gd for the same aged and fresh sample, as shown in Fig. \ref{fig:3} (d). This yields two key differences: An $0.7\pm 0.2$ eV shift of the peak at a binding energy of 141 eV, and the repression of the peak at 147 eV. Both of these changes match the earlier reports on the change of the 4d-complex of Gd upon exposure to O$_2$ \cite{WANDELT1985162,ZATSEPIN2018697}, confirming at least partial oxidation of the Gd at the Gd/TaN interface.

Considering how the results discussed above suggest that monatomic capping layers quench the Gd moment by intermixing, and TaN cap is not as effective against oxidation, we suggest using a TaN(4)/Pt(4) capping layer to passivate the magnetic properties of Gd, where the Pt layer is introduced to protect the TaN from oxidizing. Here, in Fig. \ref{fig:Figure2} (c) and (d), within the measurement window of three months we actually observe a small decrease (increase) of $\tilde{m}_\mathrm{0}$ ($T_\mathrm{Comp}$), suggesting a net increase of the Gd moment. Multilayers of Co and Gd are known to interdiffuse spontaneously,\cite{bertero_hufnagel_clemens_sinclair_1993,Andres2000} so it stands to reason that over time thermodynamically driven intermixing between the Co and Gd can be expected. Since the magnetization of Gd at these specific thicknesses occurs from the exchange interaction with the Co, a larger amount of Co neighbors will lead to a larger net moment in the Gd magnetic system. We conjecture that this has also occurred in the samples with the other capping layers, but was offset by the dominant effect of simultaneous quenching of the Gd magnetization due to intermixing between Gd and the capping layer. 

In conclusion, we have demonstrated that care has to be taken when considering layered synthetic ferrimagnets based on Gd for applications in spintronics. Significant aging effects were observed in the compensation temperature and the net moment of Co/Gd bilayer samples, which suggest a gradual quenching of the induced Gd magnetization over time. The aging effects driven by gradual intermixing described here are expected to not be restricted to the bilayer samples investigated in this work, but also to more complex synthetic ferrimagnetic multilayers based on Pt/Co/Gd-multilayers which have recently garnered attention from the research community. Although we expect no direct effect on the fundamental properties of the layered systems, small changes in the net magnetic moment may be detrimental for applications which require precise control over the magnetization. Ultimately, this work provides important guidelines for stack design involving Gd-terminated Co/Gd-based synthetic ferrimagnets, and demonstrates that with proper care of the capping layer, their magnetic properties can be effectively passivated.

\section*{Supplementary Material}
In Sup. Mat. 1, details about the sputter deposition process are presented. Sup. Mat. 2 shows a repeat study of the key experiment in the main text. In Sup. Mat. 3, experimental details regarding the EDX/STEM characterization can be found. In Sup. Mat. 4, a full range version of the EDX profile in Fig. \ref{fig:3}(b) is given. Finally, in Sup. Mat. 5, experimental details of the XPS characterization are presented.

\begin{acknowledgments}

This work was part of the research program Foundation for Fundamental Research on Matter (FOM) and Gravitation program “Research Center for Integrated Nanophotonics,” which are financed by the Dutch Research Council (NWO). This work was suported by the Eindhoven Hendrik Casimir Institute (EHCI). This project has also received funding from the European Union’s Horizon 2020 research and innovation programme under the Marie Skłodowska-Curie grant agreement No. 860060. Solliance and the Dutch province of Noord Brabant are acknowledged for funding the TEM facility. Peter Graat (Eurofins Materials Science Netherlands) is gratefully acknowledged for discussions on the TEM-EDX quantification. 

\end{acknowledgments}

\section*{Author Delcarations}

\subsection*{Conflict of Interest}
    The authors have no conflicts to disclose

\subsection*{Author Contriubtions}
\noindent
\textbf{Thomas J. Kools:} Conceptualization (equal); Data Curation (lead);
Formal analysis (lead); Investigation (equal); Validation (lead); Visualization (lead); Writing - original draft (lead); Writing - review \& editing (lead). \\
\textbf{Youri L.W. van Hees:} Conceptualization (equal); Data curation (supporting); Formal analysis (supporting); Investigation (supporting); Supervision (supporting); Validation (supporting); Writing - review \& editing (supporting). \\
\textbf{Kenneth Poissonnier} Conceptualization (supporting); Data curation (supporting); Formal analysis (supporting); Investigation (equal); Validation (supporting); Writing - review \& editing (supporting). \\
\textbf{Pingzhi Li} Conceptualization (supporting); Formal analysis (supporting); Investigation (supporting); Methodology (supporting); Writing - review \& editing (supporting).\\
\textbf{Beatriz Barcones Campo} Methodology (supporting); Writing review \& editing (supporting)\\
\textbf{Marcel A. Verheijen} Formal analysis (supporting); Investigation (supporting); Methodology (supporting); Writing – review
and editing (supporting).\\
\textbf{Bert Koopmans:} Conceptualization (equal);
Funding acquisition (equal); Supervision (equal); Writing – review
and editing (supporting).\\ 
\textbf{Reinoud Lavrijsen:} Conceptualization
(equal); Funding acquisition (equal); Supervision (equal); Writing –
review and editing (supporting).\\
    
\section*{Data Availability}
The data that support the findings of this study are available from the corresponding author upon reasonable request.

\clearpage

%% SUPPLEMENTARY ABOUT TEM/EDX CHARACTERIZATION %%

\section{Sputter deposition details}
In this section, we provide a detailed description of the deposition system and methodology that was used to fabricate the samples in this work. For all the deposited samples, a thermally oxidized Si(\SI{500}{\micro \metre})/SiO(\SI{100}{\nano \metre}) substrate was used. Prior and after to deposition, the substrate was subsequently cleaned in acetone and isopropanol in a sonicating bath and then dried immediately using a nitrogen blower.

All magnetron sputter depositions are performed in a custom-made Bestec vacuum chamber with a typical base pressure of (5$\pm4)\times$10$^{-9}$ mBar (depending on the time since venting, and prior depositions). For the plasma, we use 6.0 Ar gas, with $\leq$0.5 ppm O$_2$, H$_2$O, N$_2$ and H$_2$ impurities, and $\leq$0.1 ppm CO, CO$_2$ and C$_\mathrm{x}$H$_\mathrm{y}$ impurities. For the reactive sputtering process, 5.0 N$_2$ is used with $\leq$0.5 ppm O$_2$, and H$_2$O, and $\leq$0.2 ppm C$_\mathrm{x}$H$_\mathrm{y}$ impurities. 

The sputter targets used in this work are commercially available. The target material, target purity, $\alpha$, and target diameter, $D$, used in the depositions are listed in Table \ref{tab:table1}. Note that the TaN films were grown from a pure Ta target in a mixed Ar/N$_2$ environment at the substrate position. Before all depositions, targets were pre-sputtered for $\sim$ 5 to 10 minutes to remove any native oxide that may have formed on it, as especially Gd is known to be very susceptible to oxidation, even in the ultra-high vacuum conditions of our sputter system. During the deposition, the sample is rotated at a constant angular velocity of 10 revolutions per minute in order to promote sample homogeneity.

Table \ref{tab:table1} also lists the dissipated plasma power, $P$, pressure in the chamber $p_\mathrm{gas}$, sample-to-target distance $d$, Ar mass flow rate, $\mu_\mathrm{Ar}$, N$_\mathrm{2}$ mass flow rate, $\mu_\mathrm{N_2}$, and the growth rate $G$. These are the characteristic parameters describing the deposition processes of the Ta, Pt, Co, Gd and TaN layers. All samples discussed in the main text were grown successively, in order to minimize effects of changes in background pressure between depositions.

\begin{table*}
\caption{\label{tab:table1}Summary of the sputter target and process details for the depositions performed in this work. Listed are the target purity, $\alpha$, target diameter $D$, dissipated plasma power, $P$, pressure in the chamber $p_\mathrm{gas}$, sample-to-target distance $d$, Ar mass flow rate, $\mu_\mathrm{Ar}$, N$_\mathrm{2}$ mass flow rate, $\mu_\mathrm{N_2}$, and the growth rate $G$.}
\begin{ruledtabular}
\begin{tabular}{ccc|cccccc}
\multicolumn{3}{c}{Sputter targets}&\multicolumn{6}{c}{Sputter process}\\
Material&$\alpha$(\%)&$D$ (inch)&$P$ (W)&$p_\mathrm{gas}$ (mBar)&$d$ (mm)
&$\mu_\mathrm{Ar}$ (sccm)&$\mu_\mathrm{N_2}$ (sccm)&$G$ (\AA/s)\\ \hline
Ta&99.99&3&20&1$\times$10$^{-2}$&130&50&0&$0.56\pm0.05$ \\
Pt&99.99&2&20&1$\times$10$^{-2}$&121&50&0&$1.1\pm0.1$ \\
Co&99.99&3&20&1$\times$10$^{-2}$&130&50&0&$0.41\pm0.04$ \\
Gd&99.99&2&10&1$\times$10$^{-2}$&134&50&0&$0.65\pm0.06$ \\
 TaN\footnote{TaN is grown from the same Ta target as listed at the top of this table, but in a mixed Ar:N$_2$ environment.}&$99.99^{\text{a}}$&3$^{\text{a}}$&20&1$\times$10$^{-3}$&130&15&1.5&$0.42\pm0.04$\\

\end{tabular}
\end{ruledtabular}
\end{table*}

%% SUPPLEMENTARY ABOUT THE REPEAT STUDY%

\section{Repeat study}
\begin{figure}
\includegraphics[width=0.47\textwidth]{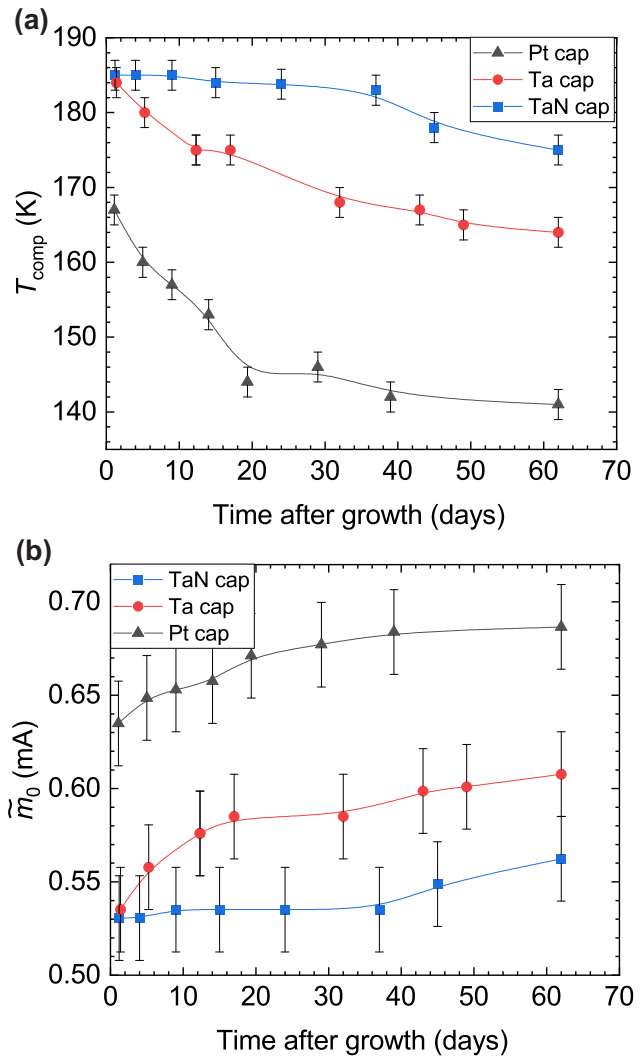}
\caption{\label{fig:repeat} Reproducibility study of stacks with the same nominal composition as in the main text. (a) Time evolution of the OOP moment, for various capping layers. (b) Time evolution of the compensation temperature for various capping layers. Solid lines are a guide to the eye.}
\end{figure}

The data presented in the main text is a repeat study from an earlier set of samples with the same composition of Ta(4)/Pt(4)/Co(1)/Gd(3)/X, with X again the capping layer. Three separate capping layers, Ta(4), Pt(4) and TaN(4) were investigated in this initial study. Similarly as in the main text, we plot $T_\mathrm{Comp}$ and $\tilde{m}_\mathrm{0}$ as measured over time with OOP VSM-SQUID in Fig. \ref{fig:repeat}(a) and (b), respectively. The same trends as for the data presented in the main text is observed. The increase (decrease) in $\tilde{m}_\mathrm{0}$ ($T_\mathrm{Comp}$) for the monatomic capping layers suggest the same quenching of the Gd magnetization as discussed in the main text. Also, similar to the data presented in the main text, the TaN initially seems to passivate the magnetic moment quite well, but eventually also starts to decay.

An important difference here is that the initial magnetic moment ($T_\mathrm{Comp}$) of the TaN is higher (lower) than that of the Ta, in contrast to the data presented in the main text. We argue that this is due to the fact that the data presented in this supplementary are obtained from samples that were not grown at identical base pressures, namely $4\times10^{-9}$ mBar for the monatomic caps and $8\times10^{-9}$ for the TaN cap. This is in contrast to the results presented in the main text, where successive depositions ensured identical deposition chamber conditions for the different samples. In general, we observe that a higher base pressure has a quenching effect on the resulting Gd moment, which we speculate is due to enhanced oxidation during the deposition process. Nonetheless, a systematic study needs to be performed to draw definitive conclusions, which is beyond the scope of this manuscript.

%% SUPPLEMENTARY ABOUT TEM/EDX CHARACTERIZATION %%

\section{TEM and EDX characterization details}
Transmission electron microscope (TEM) images were obtained with a commercially available JEOL ARM 200F TEM, equipped with a 100 mm$^2$ Centurio SDD Energy Dispersive X-ray (EDX) detector, operated at 200 kV.

From the bright field TEM, image contrast is obtained by interaction of the electron beam with the sample. Several contrast effects play a role. In the resulting TEM image, denser areas and areas containing heavier elements appear darker due to scattering of the electrons in the sample. In addition, scattering from crystal planes introduces diffraction contrast. This contrast depends on the orientation of a crystalline area in the sample with respect to the direction of the incoming electron beam. As a result, in a TEM image of a sample consisting of randomly oriented crystals, each crystal will have its own gray-level. In this way, one can distinguish between different materials, as well as image individual crystals and crystal defects.

Using an Energy Dispersive X-ray Spectroscopy (EDX) detector, it is possible to detect element characteristic X-rays. In the EDX spectrum, the detected signal is plotted as a function of the (characteristic) energy. Chemical compositions can be obtained by quantification of the data. Fig.3b displays elemental profiles after quantification of an EDX mapping of the area displayed in Fig.3a, averaged over the entire width of the image. However, quantification of signals coming from light elements (like O and N) does not necessarily lead to the right concentrations, since these signals may be absorbed by the relatively thick TEM-sample. Additionally, the carbon content is over-represented, as carbon is being deposited on the TEM sample due to cracking of hydrocarbons upon electron beam irradiation during the EDX mapping acquisition. The raw EDX spectra show Fe and Co peaks at every position in the mapping, due to spurious X-rays originating from scattering of the electron beam at the objective lens. The ratio of the Fe-K and Co-K lines of this spurious signal is 1:1. As this effect would overestimate the presence of Co throughout the stack, the Co signal as presented in Fig.3a was corrected by subtracting the Fe profile counts from the Co profile. 

TEM and EDX measurements were performed by Marcel Verheijen. Focused ion beam lift-out preparation was done by Beatriz Barcones Campo. Prior to the FIB milling, a protective SiOx(C) layer was deposited on the sample using electron beam induced deposition in the FIB. This layer is recognizable as the bright top layer in Fig.3a.

\section{Full-range EDX-measurement}

\begin{figure*}
\includegraphics[width=0.75  \textwidth]{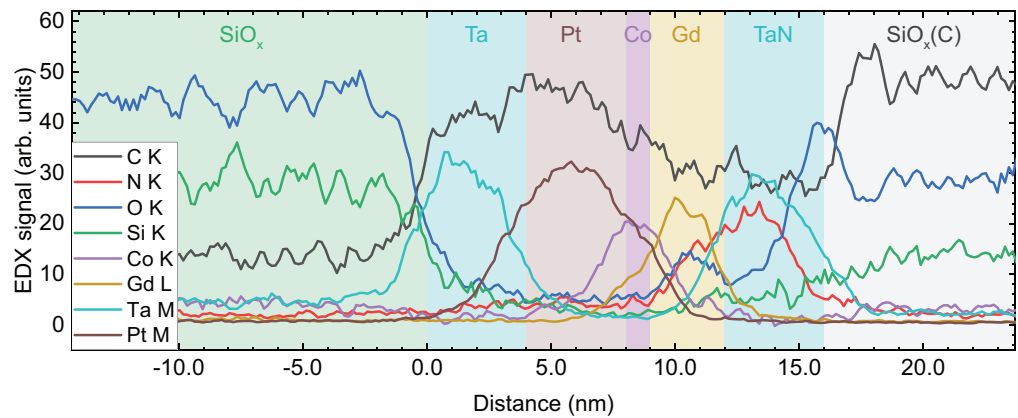}
\caption{\label{Supfig:5}  Full range elemental profile obtained by EDX of the sample investigated in figures 3(a) and 3(b) of the main text. Relative peak intensities do not represent the atomic percentages. Shaded areas indicate the nominal layer thicknesses and serve as a guide to the eye.}
\end{figure*}

Fig. \ref{Supfig:5} shows the full range plot of Fig. \ref{fig:3}(b), including the concentration of C to the spectrum.

%% SUPPLEMENTARY ABOUT XPS CHARACTERIZATION %%

\section{XPS characterization details}
\begin{table}
\caption{\label{tab:tableXPS} Overview of the instrument parameters for all presented XPS spectra.}
\begin{ruledtabular}
\begin{tabular}{cc}
Instrument property& Value/comment\\ \hline
Manufacturer and Model& Specs custom-made\\
Analyzer type& Spherical sector\\
Detector & Specs multichannel analyzer (16 channels)\\
Analyzer Mode& Constant pass energy \\
Excitation Source & Al K$_\mathrm{\alpha}$ (1486.6 eV)\\
Source Beam size & $\geq 2$ cm $\times \geq 2$ cm \\
\end{tabular}
\end{ruledtabular}
\end{table}

XPS characterization is performed in a custom-built Bestec vacuum chamber at a typical base pressure of $5\times10^{-10}$ mBar. The NanoAccess vacuum cluster at Eindhoven University of Technology allows for the transfer of samples from the sputtering chamber where the sample is fabricated to the analysis chamber where XPS is performed without breaking the ultra-high vacuum conditions. The details of the instruments used for the XPS analysis are listed in Table \ref{tab:tableXPS}.

\bibliographystyle{ieeetr}
\bibliography{ms.bib}% Produces the bibliography via BibTeXspectrum,

\newpage

\clearpage

\end{document}